\documentstyle[prd,preprint,tighten,aps,eqsecnum,amssymb,newlfont,epsfig]{revtex}
\begin{document}
\draft
\preprint{
\begin{tabular}{r}
IASSNS-AST 97/63\\
UWThPh-1997-43\\
DFTT 67/97\\
hep-ph/9711311
\end{tabular}
}
\title{Neutrino mixing from neutrino oscillation data}
\author{S.M. Bilenky}
\address{Joint Institute for Nuclear Research, Dubna, Russia, and\\
Institute for Advanced Study, Princeton, N.J. 08540}
\author{C. Giunti}
\address{INFN, Sezione di Torino, and Dipartimento di Fisica Teorica,
Universit\`a di Torino,\\
Via P. Giuria 1, I--10125 Torino, Italy}
\author{W. Grimus}
\address{Institute for Theoretical Physics, University of Vienna,\\
Boltzmanngasse 5, A--1090 Vienna, Austria}
\maketitle
\begin{abstract}
We show that the existing neutrino oscillation data
are in favour of the schemes with mixing of four massive neutrinos
and
that only two of these schemes,
with two pairs of neutrinos with close 
masses separated by a gap of about 1 eV,
are compatible with all data.
Possible implications of these schemes for future experiments are
discussed.
\end{abstract}

\pacs{Talk presented by S.M. Bilenky at the
\emph{Erice School on Nuclear Physics, $19^{\mathrm{th}}$ course
"Neutrinos in Astro, Particle and Nuclear Physics"},
16--24 September 1997.}

\section{Introduction}

The problem of neutrino masses and mixing is the key problem
of today's neutrino physics.
The investigation of this problem is
one of the major ways to search for new physics beyond the standard model.
There exist three indications in favour of non-zero neutrino masses
and neutrino mixing at present. 
The first indication comes from solar neutrino experiments
\cite{sun-exp}.
The observed deficit of solar $\nu_e$'s can be explained
by neutrino mixing with
$ \Delta{m}^2 \sim 10^{-5} \, \mathrm{eV}^2 $
in the case
of MSW transitions or with
$ \Delta{m}^2 \sim 10^{-10} \, \mathrm{eV}^2 $
in the case
of vacuum oscillations
($\Delta{m}^2$ is
the relevant neutrino mass-squared difference).
The second indication in favour of neutrino mixing
comes from the observed deficit of
atmospheric $\nu_\mu$
\cite{Kam-SK-atm,IMB-Soudan}.
From the data of the Kamiokande and 
Super-Kamiokande experiments \cite{Kam-SK-atm}
it follows that a second scale 
$ \Delta{m}^2 \sim 10^{-3}-10^{-2} \, \mathrm{eV}^2 $
exists.
The third indication
was obtained in the LSND experiment \cite{LSND}.
The data of this experiment are in favour of the existence
of a third scale
$ \Delta{m}^2 \sim 1 \, \mathrm{eV}^2 $.
There exist also data of numerous reactor and accelerator
short-baseline (SBL) experiments in which no indications
in favour of neutrino mixing were obtained (see \cite{Boehm-Vannucci}).

Which information about the neutrino mass spectrum and
neutrino mixing and
what implications for future experiments can be inferred
from the existing data?
We will discuss here these questions  
in the framework of a general phenomenological approach
that takes into account only the unitarity of the mixing matrix
\cite{BBGK95,BBGK96,BGKP,BGG96,BGG97}.

We will start with a few general remarks about neutrino masses
and mixing.
There are two possibilities for massive 
neutrinos (see \cite{BP78,BP87}):
\begin{enumerate}
\item
Massive neutrinos can be Dirac particles
like all other fundamental fermions.
In this case
the total lepton number
$ L = L_e + L_\mu + L_\tau $
is conserved and,
in the simplest case,
the number of massive neutrinos is equal to
the number of lepton flavours.
For neutrino mixing we have 
\begin{equation}
\nu_{{\alpha}L}
=
\sum_{i=1}^{3}
U_{{\alpha}i} \, \nu_{iL}
\qquad
(\alpha=e,\mu,\tau)
\,,
\label{01}
\end{equation}
where $U$ is the unitary mixing matrix and $\nu_i$
is the field of neutrinos with mass $m_i$.
\item
Massive neutrinos can be truly neutral
Majorana particles.
In this case the fields of neutrinos with a definite 
mass, $\chi_i$, satisfy the Majorana condition
\begin{equation}
\chi_i^c = C \, \overline{\chi}_i^T = \chi_i
\label{02}
\end{equation}
and there are no conserved
lepton numbers 
($C$ is the charge conjugation matrix).
The number of massive 
Majorana neutrinos
is equal to three
if only the left-handed neutrino fields,
which enter in the interaction,
enter in the neutrino mass term.
In general left-handed and right-handed 
neutrino fields
can enter in the neutrino mass term.
In this case the number $n$ of Majorana 
neutrinos is larger
than three and for the mixing we have
\begin{equation}
\nu_{{\alpha}L}
=
\sum_{i=1}^{n}
U_{{\alpha}i} \, \chi_{iL}
\,,
\qquad
(\nu_{aR})^c
=
\sum_{i=1}^{n}
U_{ai} \, \chi_{iL}
\,,
\label{04}
\end{equation}
where $\nu_{aR}$ is a right-handed (sterile) field.
Let us stress that LEP experiments
on the measurement of invisible width of Z-boson,
that proved that the number of flavour neutrinos is 
equal to three,
do not exclude the possibility
of existence of more than three massive light neutrinos
(the right-handed fields do not enter in the standard 
neutral current).
\end{enumerate}

Dirac neutrino masses can be generated by the standard
mechanism with a Higgs doublet
if the right-handed fields $\nu_{aR}$
enter in the Lagrangian and are singlets.
Majorana neutrino masses can be generated only in the framework of
theories in which lepton numbers are not conserved.
The most popular mechanism of generation of Majorana neutrino 
masses is the see-saw mechanism \cite{see-saw}.
This mechanism is based on 
the assumption that the lepton number is violated 
by the right-handed Majorana mass term
at a scale $M$ that 
is much larger than the scales of the masses of leptons and quarks 
(usually $ M \simeq M_{\mathrm{GUT}} $).
In the standard see-saw case,
for the neutrino masses we have
\begin{equation}
m_i
\simeq
\frac{ (m_{Fi})^2 }{ M }
\,.
\label{05}
\end{equation}
where $m_{Fi}$
is the mass of the lepton or up-quark in the $i^{\mathrm{th}}$-generation.
Thus, if the neutrino masses are of see-saw origin,
there are three massive neutrinos
with masses that satisfy the hierarchy relation 
\begin{equation}
m_1 \ll m_2 \ll m_3
\,.
\label{06}
\end{equation}

\section{Three massive neutrinos}

We will start with the case of three massive neutrinos
\cite{three,BBGK95,BBGK96}
and we will assume that there is a hierarchy of neutrino 
masses and 
$ \Delta{m}^2_{21} \equiv m_2^2 - m_1^2 $
is relevant for the suppression of 
the flux of solar $\nu_e$'s.
Let us consider SBL neutrino oscillations in such a scheme.
Taking into account that
$ \frac{ \Delta{m}^2_{21} L }{ 2 p } \ll 1 $
($L$ is the source-detector distance and
$p$ is the neutrino momentum), 
for the probability of
$\nu_\alpha\to\nu_\beta$
transition
we have
\begin{equation}
P_{\nu_\alpha\to\nu_\beta}
=
\left|
\delta_{\alpha\beta}
+
U_{\beta3} \, U_{\alpha3}^*
\left(
e^{ -i \frac{ \Delta{m}^2 L }{ 2 p } } - 1
\right)
\right|^2
\,,
\label{07}
\end{equation}
with
$ \Delta{m}^2 \equiv \Delta{m}^2_{31} \equiv m_3^2 - m_1^2 $.
From this expression we easily find
\begin{eqnarray}
P_{\nu_\alpha\to\nu_\beta}
\null & \null = \null & \null
\frac{1}{2}
\,
A_{\beta;\alpha}
\left( 1 - \cos\frac{ \Delta{m}^2 L }{ 2 p } \right)
\,,
\qquad \mbox{for} \qquad
\beta\neq\alpha
\,,
\label{08}
\\
P_{\nu_\alpha\to\nu_\alpha}
\null & \null = \null & \null
1
-
\frac{1}{2}
\,
B_{\alpha;\alpha}
\left( 1 - \cos\frac{ \Delta{m}^2 L }{ 2 p } \right)
\,.
\label{09}
\end{eqnarray}
with the oscillation amplitudes
$A_{\beta;\alpha}$
and
$B_{\alpha;\alpha}$
given by
\begin{eqnarray}
A_{\beta;\alpha}
\null & \null = \null & \null
4 \, |U_{\beta3}|^2 \, |U_{\alpha3}|^2
\,,
\label{10}
\\
B_{\alpha;\alpha}
\null & \null = \null & \null
\sum_{\beta\neq\alpha} A_{\beta;\alpha}
=
4 \, |U_{\alpha3}|^2 \left( 1 - |U_{\alpha3}|^2 \right)
\,.
\label{11}
\end{eqnarray}

In the case of a hierarchy of neutrino masses,
only one oscillation length characterizes neutrino oscillations in SBL experiments.
It is obvious that
the dependence of the transition probabilities on the quantity
$ \Delta{m}^2 L / 2 p $
has the same form as in the two-neutrino case.
Let us stress, however, that the expressions (\ref{08}) and (\ref{09})
are general and describe 
transitions between all three flavour neutrinos.
Notice also that in the case of a hierarchy of neutrino masses
the CP phase does not enter in the expressions for the transition 
probabilities.
As a result we have
\begin{equation}
P_{\nu_\alpha\to\nu_\beta}
=
P_{\bar\nu_\alpha\to\bar\nu_\beta}
\,.
\label{12}
\end{equation}

With the help of Eqs.(\ref{09}) and (\ref{11}),
one can obtain bounds
on the mixing parameters
$|U_{e3}|^2$ and $|U_{\mu3}|^2$
from exclusive plots that were found from the data
of SBL reactor and accelerator disappearance experiments 
(because of unitarity of 
the mixing matrix
$ |U_{\tau3}|^2 = 1 - |U_{e3}|^2 - |U_{\mu3}|^2 $).

We will consider the range
\begin{equation}
10^{-1} \, \mathrm{eV}^2
\leq
\Delta{m}^2
\leq
10^{3} \, \mathrm{eV}^2
\,.
\label{13}
\end{equation}
From the exclusion curves
of SBL disappearance experiments,
at any fixed value of
$\Delta{m}^2$
we obtain the upper bounds 
$ B_{\alpha;\alpha} \leq B_{\alpha;\alpha}^0 $
for $\alpha=e,\mu$.
From Eq.(\ref{11}),
for the mixing parameters
$|U_{\alpha3}|^2$
we have
\begin{equation}
|U_{\alpha3}|^2 \leq a_{\alpha}^0
\qquad \mbox{or} \qquad
|U_{\alpha3}|^2 \geq 1 - a_{\alpha}^0
\,,
\qquad \mbox{with} \qquad
a_{\alpha}^0
=
\frac{1}{2}
\left( 1 - \sqrt{ 1 - B_{\alpha;\alpha}^0 } \right)
\,.
\label{14}
\end{equation}
We have obtained the values of $a_{e}^0$ and $a_{\mu}^0$
from the exclusion plots of 
the Bugey reactor experiment \cite{Bugey95}
and the CDHS and CCFR
accelerator experiments
\cite{CDHS84-CCFR84}
(see Fig.1 of \cite{BBGK96}).
In the  range (\ref{13})  
we have
$ a_{e}^0 \lesssim 4 \times10^{-2} $
and
$ a_{\mu}^0 \lesssim 2 \times 10^{-1} $
(for $ \Delta{m}^2 \gtrsim 0.3 \, \mathrm{eV}^2 $).
Thus, from the results of disappearance experiments
it follows that the mixing parameters
$|U_{e3}|^2$
and
$|U_{\mu3}|^2$
can be either small or large (close to one).

Now let us take into account the results of solar neutrino 
experiments.
For the probability of solar neutrinos to survive in the case
of neutrino mass hierarchy we have the following lower bound:
\begin{equation}
P_{\nu_e\to\nu_e}^{\mathrm{sun}}
\geq
|U_{e3}|^4
\,.
\label{16}
\end{equation}
If
$ |U_{e3}|^2 \geq 1 - a_{e}^0 $,
from (\ref{16})
it follows that at all solar neutrino energies
$ P_{\nu_e\to\nu_e}^{\mathrm{sun}} \gtrsim 0.92 $.
This is not compatible with the results of solar neutrino
experiments.
Thus, the mixing parameter
$|U_{e3}|^2$
must be small:
$ |U_{e3}|^2 \leq a_{e}^0 $.

We come to the conclusion that from the results 
of SBL inclusive experiments and solar neutrino
experiments it follows that in the case of three massive neutrinos
with a hierarchy of masses 
only two schemes of mixing are possible:
\begin{equation}
\mathrm{I.}
\left\{
\begin{array}{l} \displaystyle
|U_{e3}|^2 \leq a_{e}^0
\,,
\\ \displaystyle
|U_{\mu3}|^2 \leq a_{\mu}^0
\,,
\end{array}
\right.
\qquad \qquad
\mathrm{II.}
\left\{
\begin{array}{l} \displaystyle
|U_{e3}|^2 \leq a_{e}^0
\,,
\\ \displaystyle
|U_{\mu3}|^2 \geq 1 - a_{\mu}^0
\,.
\end{array}
\right.
\label{17}
\end{equation}

Let us consider
$\nu_\mu\leftrightarrows\nu_e$
oscillations
in the case of scheme I.
From Eqs.(\ref{10}) and (\ref{17}),
for the oscillation amplitude we have
\begin{equation}
A_{e;\mu}
\leq
4 \, |U_{e3}|^2 \, |U_{\mu3}|^2
\leq
4 \, a_{e}^0 \, a_{\mu}^0
\,,
\label{18}
\end{equation}
i.e. the upper bound for the amplitude $A_{e;\mu}$
is quadratic in the small quantities  
$a_{e}^0$, $a_{\mu}^0$;
thus,
in the case of scheme I
$\nu_\mu\leftrightarrows\nu_e$
oscillations are strongly suppressed.

Let us compare now the upper bound (\ref{18})
with the results of the LSND experiment
in which
$\nu_\mu\leftrightarrows\nu_e$
oscillations were observed.
In Fig.\ref{fig1}
the shadowed region in the
$A_{e;\mu}$--$\Delta{m}^2$ plane
is the region allowed by the results of the LSND experiment. 
The regions excluded by the Bugey experiment \cite{Bugey95}
and by the BNL E734, BNL E776 and CCFR experiments
\cite{BNLE734-BNLE776-CCFR96}
are also shown.
The upper bound (\ref{18}) is presented by the curve passing through
the circles.
As it is seen from Fig.\ref{fig1},
the upper bound (\ref{18}) is not
compatible with the results of the LSND experiment if the results of other
oscillation experiments are taken into account.
Thus, the scheme I with a hierarchy of masses and couplings,
similar to the hierarchy that takes place in the quark sector,
is not favoured by the results of SBL experiments.

In the case of scheme II,
the upper bound of the amplitude $A_{e;\mu}$
is linear in the small quantity $a_e^0$:
$ A_{e;\mu} \leq 4 a_e^0 $.
This upper bound is compatible with the LSND data.
Note that, if scheme II is realized in nature,
$ | \nu_\mu \rangle \simeq | \nu_3 \rangle $
and the vectors
$ | \nu_e \rangle $
and
$ | \nu_\tau \rangle $
are superpositions of
$ | \nu_1 \rangle $
and
$ | \nu_2 \rangle $.
For the neutrino masses we have 
$ m_{\nu_\mu} \simeq m_3 $,
$ m_{\nu_e} , m_{\nu_\tau} \ll m_3 $.

Up to now we did not consider atmospheric neutrinos.
In the framework
of the scheme with three massive neutrinos and a neutrino mass 
hierarchy there are only two possibilities to take into account 
the atmospheric neutrino anomaly:
\begin{enumerate}
\item
To assume that
$\Delta{m}^2_{21}$
is relevant for the suppression of solar 
$\nu_e$'s and for the atmospheric neutrino anomaly
\cite{AP97}.
\item
To assume that
$\Delta{m}^2_{31}$
is relevant for the LSND anomaly and 
for the atmospheric neutrino anomaly \cite{CF97}.
\end{enumerate}

In first case,
the average survival probability of solar $\nu_e$'s is constant 
(this is disfavoured
by the data of
solar neutrino experiments \cite{KP97,CMMV97})
and the parameters
$|U_{e3}|^2$, $|U_{\mu3}|^2$
satisfy inequalities
$ |U_{e3}|^2 \leq a_e^0 $ and $ |U_{\mu3}|^2 \leq a_\mu^0 $
(that are not compatible with the LSND 
result, as we have discussed above).

In the second case
it is not possible to explain
the angular dependence of the double ratio of muon and electron events
that was observed by the Kamiokande and Super-Kamiokande experiments 
\cite{Kam-SK-atm}.

All the existing indications in favour of neutrino mixing will be
checked by several experiments that now are under preparation.
If for the time being we accept them,
we come to the necessity of consideration of schemes
with four massive neutrinos, that include the three flavour  
neutrinos $\nu_e$, $\nu_\mu$, $\nu_\tau$
and a sterile neutrino 
\cite{four,BGKP,BGG96,BGG97}.

\section{Four massive neutrinos}

There are six possible types of mass spectra with four neutrinos
that can accommodate 
three different scales of $\Delta{m}^2$.
Let us start with the case of a hierarchy of neutrino masses
$ m_1 \ll m_2 \ll m_3 \ll m_4 $,
assuming that
$\Delta{m}^2_{21}$
is relevant for the suppression
of solar $\nu_e$'s and
$\Delta{m}^2_{31}$
is relevant for the atmospheric neutrino anomaly.
The SBL transition probabilities are given in this case by the expressions
(\ref{08})--(\ref{11})
with
$ |U_{\alpha3}|^2 \to |U_{\alpha4}|^2 $
and
$ \Delta{m}^2 \equiv \Delta{m}^2_{41} \equiv m_4^2 - m_1^2 $.
From SBL inclusive data in the range (\ref{13}) of
$\Delta{m}^2$,
we have
\begin{equation}
|U_{\alpha4}|^2 \leq a_{\alpha}^0
\qquad \mbox{or} \qquad
|U_{\alpha4}|^2 \geq 1 - a_{\alpha}^0
\,,
\qquad \mbox{for} \qquad
\alpha=e,\mu
\,.
\label{19}
\end{equation}

For the survival probability of the atmospheric $\nu_\mu$'s
in the scheme under consideration we have the lower 
bound \cite{BGG96}
\begin{equation}
P_{\nu_\mu\to\nu_\mu}^{\mathrm{atm}} \geq |U_{\mu4}|^4
\,.
\label{20}
\end{equation}
Now, from Eq.(\ref{16}) with
$ |U_{e3}|^2 \to |U_{e4}|^2 $
and from Eq.(\ref{20}) it follows that
large values of the mixing parameters
$|U_{e4}|^2$ and $|U_{\mu4}|^2$
are not compatible with solar and atmospheric neutrino data.
We come to the conclusion that both mixing 
parameters
$|U_{e4}|^2$ and $|U_{\mu4}|^2$
are small:
$ |U_{e4}|^2 \leq a_e^0 $ and $ |U_{\mu4}|^2 \leq a_\mu^0 $.
As in the case of scheme I for three neutrinos,
in the scheme under consideration
the SBL amplitude
$A_{e;\mu}$
is constrained by the upper bound (\ref{18})
(with $ |U_{\alpha3}|^2 \to |U_{\alpha4}|^2 $),
which is not compatible with the LSND allowed region 
(see Fig.\ref{fig1}). 
Thus, 
a mass hierarchy of four neutrinos is not favoured by the existing
data. The same conclusion can be drawn for all four-neutrino mass
spectra with one neutrino mass separated from the group of three
close masses by the ``LSND gap'' ($\sim 1$ eV). 
 
Let us consider now the two possible neutrino mass spectra
\begin{equation}
(\mathrm{A})
\qquad
\underbrace{
\overbrace{m_1 < m_2}^{\mathrm{atm}}
\ll
\overbrace{m_3 < m_4}^{\mathrm{sun}}
}_{\mathrm{LSND}}
\,,
\qquad \qquad
(\mathrm{B})
\qquad
\underbrace{
\overbrace{m_1 < m_2}^{\mathrm{sun}}
\ll
\overbrace{m_3 < m_4}^{\mathrm{atm}}
}_{\mathrm{LSND}}
\,,
\label{21}
\end{equation}
with two groups of close masses separated by a $\sim 1$ eV gap.
In the case of such neutrino mass spectra,
the SBL transition probabilities are given by the expressions
(\ref{08}) and (\ref{09})
and the
oscillation amplitudes are given by
\begin{equation}
A_{\beta;\alpha}
=
4 \left| \sum_i U_{{\beta}i} \, U_{{\alpha}i}^{*} \right|^2
\,,
\qquad
B_{\alpha;\alpha}
=
4
\left( \sum_i |U_{{\alpha}i}|^2 \right)
\left( 1 - \sum_i |U_{{\alpha}i}|^2 \right)
\,.
\label{22}
\end{equation}
where the index $i$ runs over $1,2$ or $3,4$.
From the exclusion plots of the Bugey, CDHS and CCFR
disappearance experiments we have
\begin{equation}
\sum_i |U_{{\alpha}i}|^2 \leq a_{\alpha}^0
\qquad \mbox{or} \qquad
\sum_i |U_{{\alpha}i}|^2 \geq 1 - a_{\alpha}^0
\,,
\qquad \mbox{for} \qquad
\alpha=e,\mu
\,.
\label{23}
\end{equation}

If we take into account the results of solar and atmospheric 
neutrino experiments, in the case of scheme A we have
\begin{equation}
\sum_{i=1,2} |U_{ei}|^2 \leq a_e^0
\qquad \mbox{and} \qquad
\sum_{i=3,4} |U_{{\mu}i}|^2 \leq a_\mu^0
\,.
\label{24}
\end{equation}
The corresponding inequalities in the scheme B can be obtained
from Eq.(\ref{24}) with the change
$ 1,2 \leftrightarrows 3,4 $.

Now, for the amplitude of
$\nu_\mu\leftrightarrows\nu_e$
oscillations, 
from Eqs.(\ref{22}) and (\ref{24}),
in both schemes we have the upper bound
\begin{equation}
A_{e;\mu}
=
4 \left| \sum_i U_{{\mu}i} \, U_{ei}^{*} \right|^2
\leq
4
\left( \sum_i |U_{{\mu}i}|^2 \right)
\left( \sum_i |U_{ei}|^2 \right)
\leq
4 \, \mathrm{Min}[ a_e^0 , a_\mu^0 ]
\,,
\label{25}
\end{equation}
that is compatible with the LSND result.
Thus, schemes A and B can accommodate all neutrino oscillation data.

The schemes A and B 
give different predictions 
for the neutrino mass
$m(^3\mathrm{H})$
measured in
$^3\mathrm{H}$-decay
experiments, and for the effective Majorana mass
$
\langle{m}\rangle
=
\sum_{i=1}^4 U_{ei}^2 m_i
$
that enters into matrix element of neutrinoless double-beta decay.
In the scheme A we have
\begin{equation}
m(^3\mathrm{H}) \simeq m_4
\,.
\label{26}
\end{equation}
In the case of scheme B,
the contribution to the beta-spectrum of the term that includes 
the heaviest masses $ m_3 \simeq m_4 $
is suppressed by the factor
$ \sum_{i=3,4} |U_{ei}|^2 \leq a_e^0 \lesssim 4 \times 10^{-2} $.

For the the effective Majorana mass
in neutrinoless double-beta decay,
in the schemes A and B we have
\begin{equation}
(\mathrm{A})
\qquad
|\langle{m}\rangle|
\leq
\sum_{i=3,4} |U_{ei}|^2 m_4
\leq
m_4
\,,
\qquad \qquad
(\mathrm{B})
\qquad
|\langle{m}\rangle|
\leq
a_e^0 m_4
\ll
m_4
\,.
\label{27}
\end{equation}

Thus, if the scheme A is realized in nature,
the $^3\mathrm{H}$-decay experiments on the 
direct measurement of neutrino mass and
the experiments on the search 
for neutrinoless double-beta decay can see
the effect of the ``LSND neutrino mass''.  

Finally,
we will consider neutrino oscillations in long-baseline (LBL) experiments
in the framework of the schemes A and B.
We will show that the data SBL experiments imply
rather strong constrains on the LBL probabilities of
$\bar\nu_e\to\bar\nu_e$
and
$\nu_\mu\to\nu_e$
transitions \cite{BGG97}.
In the scheme A,
for the probability of LBL
$\nu_\alpha\to\nu_\beta$
transitions we have the following expression:
\begin{equation}
P^{(\mathrm{LBL,A})}_{\nu_\alpha\to\nu_\beta}
=
\left|
U_{\beta1}
\,
U_{\alpha1}^{*}
+
U_{\beta2}
\,
U_{\alpha2}^{*}
\,
\exp\!\left(
- i
\frac{ \Delta{m}^{2}_{21} \, L }{ 2 \, p }
\right)
\right|^2
+
\left|
\sum_{k=3,4}
U_{{\beta}k}
\,
U_{{\alpha}k}^{*}
\right|^2
\,.
\label{28}
\end{equation}
The probability of the
$\nu_\alpha\to\nu_\beta$
transitions in scheme B
can be obtained from Eq.(\ref{28}) with the change
$ 1,2 \leftrightarrows 3,4 $.
Let us notice also that the probability of LBL
$\bar\nu_\alpha\to\bar\nu_\beta$
transition can be obtained from Eq.(\ref{28}) with the change
$ U_{{\alpha}k} \to U_{{\alpha}k}^{*} $.

Two reactor experiments,
CHOOZ \cite{CHOOZ} and Palo Verde \cite{PV},
are the first long-baseline experiments.
From Eq.(\ref{28}), for the 
probabilities of
$\bar\nu_e\to\bar\nu_e$
transition in the schemes A and B we have the 
following lower bounds:
\begin{equation}
(\mathrm{A})
\qquad
P^{(\mathrm{LBL,A})}_{\bar\nu_e\to\bar\nu_e}
\geq
\left( \sum_{i=3,4} |U_{ei}|^2 \right)^2
\,,
\qquad \qquad
(\mathrm{B})
\qquad
P^{(\mathrm{LBL,B})}_{\bar\nu_e\to\bar\nu_e}
\geq
\left( \sum_{i=1,2} |U_{ei}|^2 \right)^2
\,.
\label{29}
\end{equation}
Now, taking into account the unitarity of the mixing matrix, 
we can conclude that the quantities
in the right-hand sides of the two inequalities (\ref{29})
are large.
Thus,
on the basis of the existing neutrino oscillation data,
we expect that the probability 
$P^{(\mathrm{LBL})}_{\bar\nu_e\to\bar\nu_e}$
is close to one.
Indeed, from Eq.(\ref{29}),
for both schemes we have
\begin{equation}
P^{(\mathrm{LBL})}_{\bar\nu_e\to\bar\nu_e}
\geq
\left( 1 - a_e^0 \right)^2
\,.
\label{30}
\end{equation}
For the transition probability of $\bar\nu_e$
into any other state, from Eq.(\ref{30}) we have
the upper bound
\begin{equation}
1 - P^{(\mathrm{LBL})}_{\bar\nu_e\to\bar\nu_e}
=
\sum_{\alpha\neq{e}}
P^{(\mathrm{LBL})}_{\bar\nu_e\to\bar\nu_\alpha}
\leq
a_e^0 \left( 2 - a_e^0 \right)
\,.
\label{31}
\end{equation}
The value of $a_e^0$ depends on the SBL parameter
$\Delta{m}^2$.
In Fig.\ref{fig2} we have drawn
the curve corresponding to the upper bound (\ref{31})
for $\Delta{m}^2$
in the interval (\ref{13}). 
The minimum values of the
probability
$
\sum_{\alpha\neq{e}}
P^{(\mathrm{LBL})}_{\bar\nu_e\to\bar\nu_\alpha}
$
that is planned to be reached in the CHOOZ and Palo Verde
experiments are also shown.
The shadowed region in Fig.\ref{fig2} is the region that is allowed
(at 90\% CL)
by the data of the LSND experiment and of the other SBL experiments.
Thus, as it is seen from Fig.\ref{fig2},
in the framework of the schemes A and B,
the existing data put rather severe constraints on the LBL
transition probability of $\bar\nu_e$
into any other state.

Taking into account the unitarity of the mixing matrix and the CPT-theorem,
it easy to see that the LBL probability of
$\nu_\mu\to\nu_e$
transitions
is also strongly suppressed.
Indeed, we have
\begin{equation}
P^{(\mathrm{LBL})}_{\nu_\mu\to\nu_e}
=
P^{(\mathrm{LBL})}_{\bar\nu_e\to\bar\nu_\mu}
\leq
a_e^0 \left( 2 - a_e^0 \right)
\,.
\label{32}
\end{equation}
Another upper bound on the LBL probability of
$\nu_\mu\to\nu_e$
transitions
can be obtained from Eqs.(\ref{24}) and (\ref{28}).
For both models we have
\begin{equation}
P^{(\mathrm{LBL})}_{\nu_\mu\to\nu_e}
\leq
a_e^0 + \frac{1}{4} \, A_{e;\mu}
\,.
\label{33}
\end{equation}

The upper bound for the LBL probability of
$\nu_\mu\to\nu_e$
transitions,
obtained with the help of Eqs.(\ref{32}) and (\ref{33}),
is shown in Fig.\ref{fig3} by the short-dashed curve.
The solid line presents
the corresponding bound with matter corrections
for the K2K experiment \cite{K2K}.
The dash-dotted vertical line 
presents the expected minimal value of the probability
$P^{(\mathrm{LBL})}_{\nu_\mu\to\nu_e}$
that will be reached in the K2K experiment. 
Notice that at all the considered values of
$\Delta{m}^2$
this probability is larger 
than the upper bound with matter corrections.
The solid line in Fig.\ref{fig4} shows the bound corresponding
to Eqs.(\ref{32}) and (\ref{33})
with matter corrections
for the MINOS \cite{MINOS} and ICARUS \cite{ICARUS} experiments,
whose expected sensitivities are represented, respectively, by the
dash-dotted and dash-dot-dotted lines.
One can see that these sensitivities are sufficient to explore
the region allowed by the results of the LSND and the other SBL
experiments.

\section{Conclusions}

The existing indications in favour of neutrino oscillations   
(coming from solar, atmospheric and the LSND experiments)
require schemes with mixing of four massive neutrinos.
We have shown that only two possible schemes
with two pairs of neutrinos with close masses separated by a gap 
of about 1 eV can accommodate all the existing neutrino oscillation data.
We have discussed the possible implications
of these schemes for the experiments on the direct measurement
of the electron neutrino mass,
on the search for neutrinoless double-beta 
decay and for long-baseline neutrino experiments\footnote{
After we finished this paper the first results of the CHOOZ experiment
appeared \cite{CHOOZ97}.
No indications in favor of
$\bar\nu_e\to\bar\nu_e$
transitions were found in this experiment.
The upper bound for the
probability
$ 1 - P^{(\mathrm{LBL})}_{\bar\nu_e\to\bar\nu_e} $
found in the CHOOZ experiment is in agreement with
the bound obtained in \cite{BGG97} and presented in Fig.\ref{fig2}.
}.

\acknowledgments

S.M.B. would like to acknowledge support from
the Dyson Visiting
Professor Funds at the Institute for Advanced Study.

\newpage

\begin{minipage}[p]{0.95\textwidth}
\begin{center}
\mbox{\epsfig{file=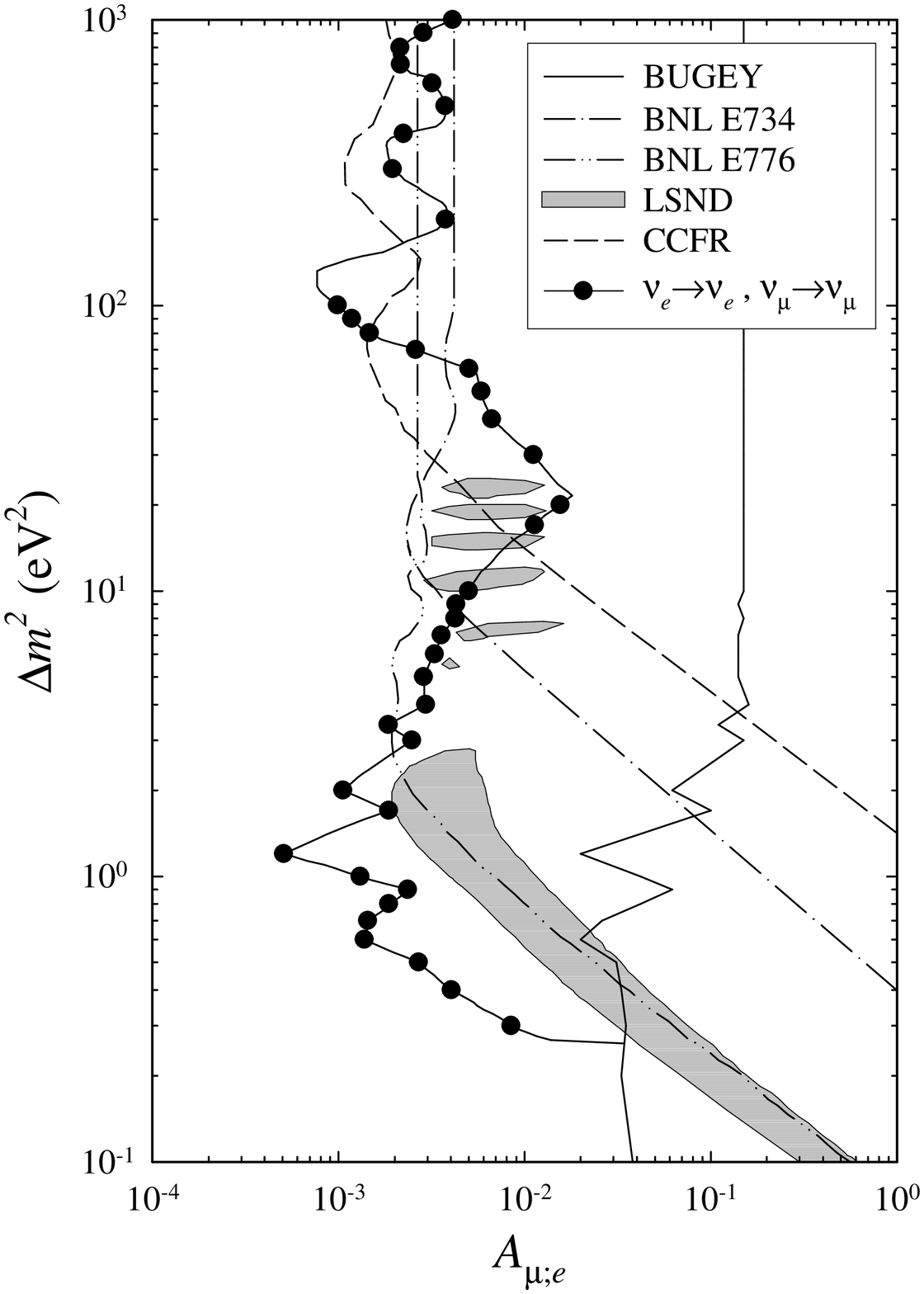,height=0.95\textheight}}
\end{center}
\end{minipage}
\begin{center}
\refstepcounter{figure}
\label{fig1}                 
\Large Figure \ref{fig1}
\end{center}

\newpage

\begin{minipage}[p]{0.95\textwidth}
\begin{center}
\mbox{\epsfig{file=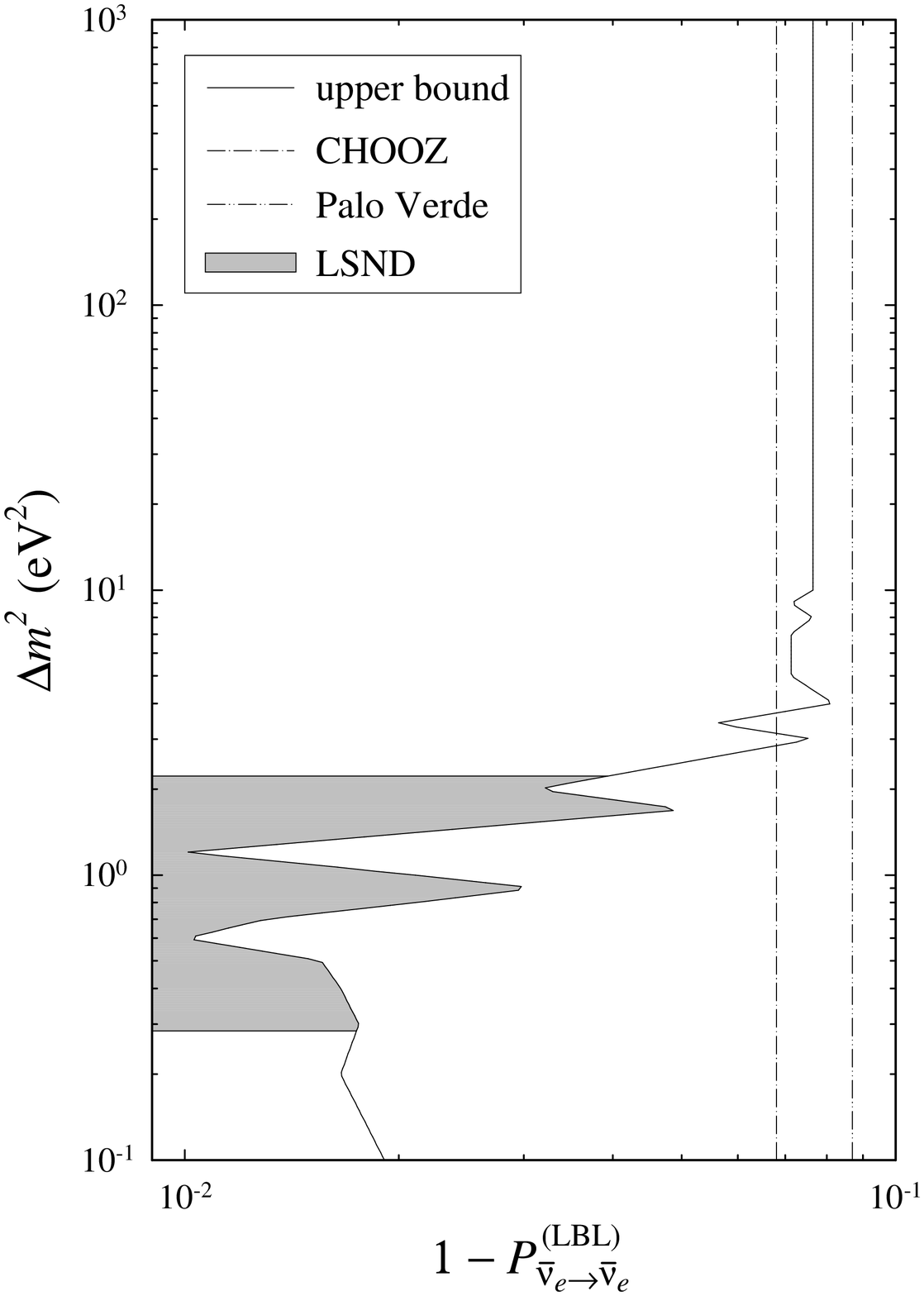,height=0.95\textheight}}
\end{center}
\end{minipage}
\begin{center}
\refstepcounter{figure}
\label{fig2}                 
\Large Figure \ref{fig2}
\end{center}

\newpage

\begin{minipage}[p]{0.95\textwidth}
\begin{center}
\mbox{\epsfig{file=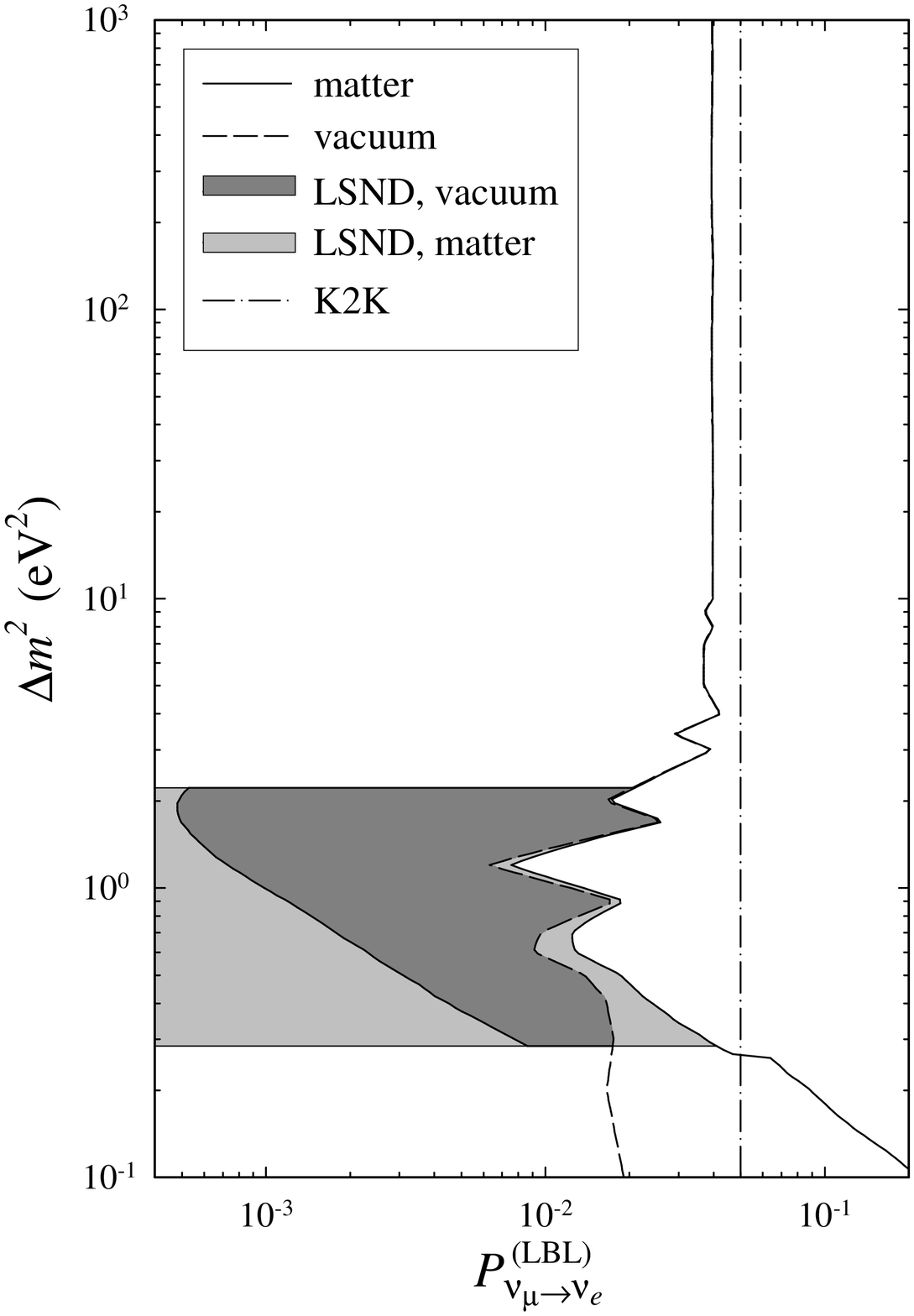,height=0.95\textheight}}
\end{center}
\end{minipage}
\begin{center}
\refstepcounter{figure}
\label{fig3}                 
\Large Figure \ref{fig3}
\end{center}

\newpage

\begin{minipage}[p]{0.95\textwidth}
\begin{center}
\mbox{\epsfig{file=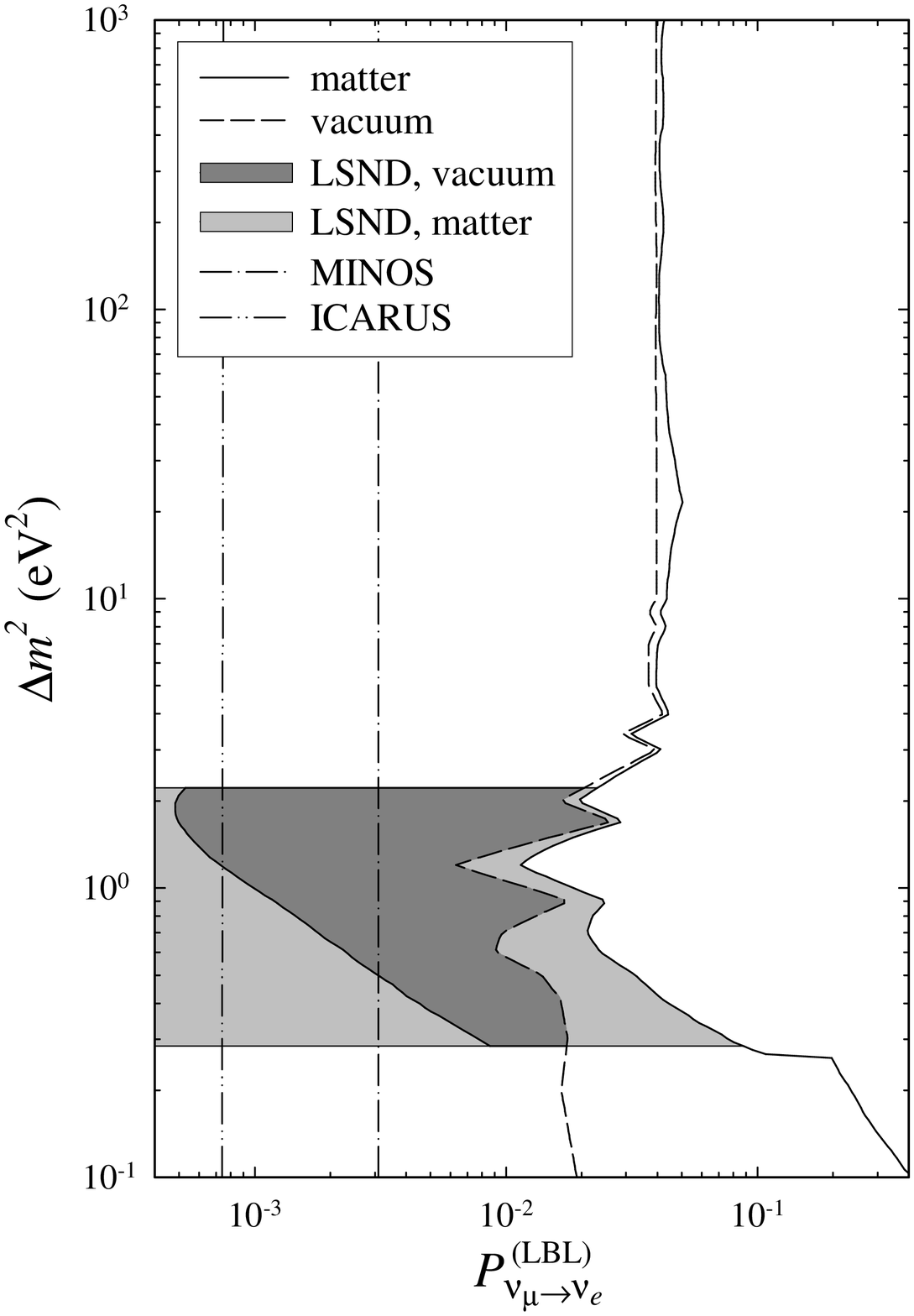,height=0.95\textheight}}
\end{center}
\end{minipage}
\begin{center}
\refstepcounter{figure}
\label{fig4}                 
\Large Figure \ref{fig4}
\end{center}

\end{document}